\documentclass[MSNbibl,nameyear,dvips]{arxstspdf}
\usepackage{flushend}
\usepackage{stfloats}

\volume{29}
\issue{2}
\pubyear{2014}
\firstpage{240}
\lastpage{241}
\doi{10.1214/14-STS470} 
\referstodoi{10.1214/13-STS457}

\makeatletter

\newcommand{\textcite}[1]{\citet{#1}}

\makeatother

\begin{document}
\begin{frontmatter}

\vspace*{6pt}\title{Discussion of ``On the Birnbaum Argument for the Strong
Likelihood Principle''}
\runtitle{Discussion}

\begin{aug}
\author[a]{\fnms{A. P.}~\snm{Dawid}\corref{}\ead[label=e1]{a.p.dawid@statslab.cam.ac.uk}}
\runauthor{A. P. Dawid}

\affiliation{Cambridge University}

\address[a]{A. P. Dawid is Emeritus Professor of Statistics,
Statistical Laboratory,
Centre for Mathematical Sciences, Cambridge University,
Wilberforce Road, Cambridge CB3 0WB, UK \printead{e1}.}
\end{aug}

%
\begin{abstract}
Deborah Mayo claims to have refuted Birnbaum's argument
that the Likelihood Principle is a logical consequence of the
Sufficiency and Conditionality Principles. However, this claim fails
because her interpretation of the Conditionality Principle is
different from
Birnbaum's. Birnbaum's proof cannot be so readily dismissed.
\end{abstract}

%
\end{frontmatter}

Deborah Mayo (\citeyear{mayoSS}) is not the first devoutly to wish
that the (strong) Likelihood Principle [principle L of
\textcite{birnbaum1962}] was \emph{not} a logical consequence of the
Sufficiency Principle (Birnbaum's S) and the Conditionality Principle
(Birnbaum's C). This concern arises because much of frequentist
inference is in clear violation of L, while at the same time
purporting to abide by S and C. This constitutes a
self-contradiction, which frequentists are, however, loth to admit.
Birnbaum himself appears to have been quite distraught at his own
finding, and in the half-century since publication of his argument
there has been a constant trickle of attempts to come to terms with
it, including one or two of my own
(\cite{apd:conform}; \cite{apd:encinf1}; \cite{apd:fraserdis}; \cite{apd:basu}); a detailed
account that I consider displays the underlying logic clearly can be
found in Chapter II ``Principles of Inference'' of \textcite{apd:pos}.

Those who feel disquiet at the destructive implications of Birnbaum's
theorem for their favored method of inference (be it frequentist or,
for example, ``objective Bayesian,'' which also violates L) have a
number of strategies to try and ease that disquiet. If they accept
the validity of the theorem, they might argue [along with
\citet{fraser1963}; \citet{durbin1970}; \citet{kalbfleisch1975}] that S or C should
not be taken as universally applicable---thus evading the consequent
of the theorem by denying its antecedents. This is at least a
logically sound ploy, although it reeks of adhockery. Also, the ploy
may not be totally successful, since some of the ``undesirable''
implications of the theorem may survive weakening of its hypotheses:
\textcite{apd:fraserdis} suggests that the principle of the
irrelevance of the stopping rule is one such survivor.

A second possible strategy is to fully accept S and C and Birnbaum's
argument---and thereby come to accept L. This is the path of
enlightenment followed by conversion.

The third strategy involves accepting S and C, but still rejecting L.
If that is your motivation (and you care about self-consistency), you
have no option but to try and find fault with the logic of Birnbaum's
theorem. This is Mayo's strategy. The only problem is that
Birnbaum's theorem is indeed logically sound. That means that Mayo's
attempt to argue the contrary must itself be unsound. Although there
are many points at which I am deeply critical of her argument, I will
content myself with drawing attention to her principal
misunderstanding, which vitiates her entire enterprise: she simply has
not grasped Birnbaum's conditionality principle C, conflating and
confusing it with Cox's WCP, which is quite different.

According to Mayo, WCP requires that ``one should condition on the
known experiment,'' or (as she phrases it in Section~4.3) ``eschew
unconditional formulations.'' But Birnbaum describes his principle C
as the requirement that
\begin{quote}
the evidential meaning of any outcome of any mixture experiment is
the same as that of the corresponding outcome of the corresponding
component experiment, ignoring the over-all structure of the mixture
experiment.
\end{quote}
That is, Birnbaum's principle C requires \emph{identity} of the
inferences to be drawn (from the same data) in different
circumstances. This imposes an equivalence relationship across such
circumstances. Principle C has nothing to say about the form or
nature of the inferences, and---importantly---unlike WCP is
entirely nondirectional. Mayo has misconstrued it as synonymous with
WCP, which would require that we should discard whatever inference we
might have been contemplating in the mixture experiment and replace
it by our favored inference in the component experiment. However, an
equally (in)valid reading of C would be the contrary: that we should
discard a contemplated component-experiment inference in favor of an
inference formed for the mixture experiment. In fact, neither of these
interpretations has anything to do with principle C and, typically---as
indeed follows from Birnbaum's theorem and the fact that
frequentist inference violates C---neither of them can be
implemented consistently within a frequentist framework.

In her Section~4.3.3 Mayo does consider the relationship between WCP and
equivalence principles, and quite correctly decides that WCP is not
one of these. In Section~7 she opines, ``The problem stems from mistaking
WCP as the equivalence\ldots.'' So at least she realises that WCP and
Birnbaum's principle C are different. However the ``problem'' is just
the contrary: she has mistaken Birnbaum's equivalence requirement C as
the ``nonequivalence'' principle WCP.

Mayo has attempted to argue that L does not follow from S and WCP.
Notwithstanding the shortfalls in her arguments, I agree with that
conclusion. The trouble is, it has nothing to do with Birnbaum's
theorem. Mayo has been attacking a straw man, and Birnbaum's result,
S \& C $\Rightarrow$ L, remains entirely untouched by her criticisms.




\end{document}